 \documentstyle[preprint,aps,pra]{revtex}

\def\bfr{{\bf r}}
\def\bfk{{\bf k}}

\def\eps{\epsilon}

\def\tl{\tilde}
\begin{document}
\title{Condensate of a charged boson fluid
 at non-integer dimensions}
\author{Sang-Hoon Kim,$^1$ Chul Koo Kim,$^{2}$ and Kyun Nahm$^3$}
\address{$^1$Division of Liberal Arts, Mokpo National Maritime University,
 Mokpo 530-729, Korea}
\address{$^2$Institute of Physics and Applied Physics,
  Yonsei University, Seoul 120-749, Korea}
\address{$^3$Department of Physics, Yonsei University, Wonju 220-710, Korea}
\date{\today}
\maketitle
\draft
\begin{abstract}
Condensate  of a charged boson fluid  at non-integer dimensions
between 2 and 3 is studied.
Interaction  between particles is assumed to be Coulombic,   
and Bogoliubov approximation is applied for a weak coupling regime.
The condensate and the superfluid fraction at finite temperatures 
and at non-integer dimensions are calculated.
The theoretical results are compared with the 
 superfluid densities of superconducting films,
which show a universal splitting behavior.
The qualitative similarity between the charged boson fluid 
and the superconducting films gives a strong clue for
 the origin of the universality.
\end{abstract} 
\newpage

\section{introduction}

Charged boson fluid(CBF) has been attracting attention
 as an interesting model
of many physical systems at low temperatures. 
Several theoretical and computational studies have been
carried out on the 2D and the 3D CBF at zero and 
finite temperatures\cite{cher,tosi1,alex2,tosi2}.
Also, a general expression for the superfluid density of dilute
Bose gas for $D > 2$ was given by Fisher and Hohenberg\cite{fish}

It has been, recently, noted that porous media and films can be studied
using fractal dimensionality between $2 < D < 3$\cite{kim4}.
This study also revealed that several prominent features of 
porous media originates from dimensionality and detailed nature
of mutual interaction plays a much less prominent role.

In this paper, we study the condensate and the superfluid density
of CBF at finite temperatures in the dimensionality between $2 < D < 3$.
Then, the result will be compared with the experimental data from
superconducting thin films. It will be shown that the nature of
gap states and detailed property of superconducting mechanism plays
a much less prominent role than that of the geometrical factors. 

We will consider a model of point-like spinless charged bosons 
embedded in a uniform neutralizing background
and coupled by a Coulomb type interaction.
The Bogoliubov approximation will be applied for 
the model Hamiltonian in weakly interacting regime to calculate
the condensate.
The condensate fraction will be calculated as a function of temperature
 and dimensionality between 2 and 3.
Then, the result is utilized to calculate the superfluid fraction 
 with the help of the
$D$-dimensional Fisher and Hohenberg formula\cite{fish}.

Finally, the theoretical results on the charged bosons in fractal dimensions
between 2 and 3 will be compared to experimental data on superfluid density
of superconducting films and the physical implications will
be discussed.

\section{Bogoliubov approach to a charged boson fluid}

We begin our theoretical scheme from the well-established Bogoliubov 
approach for the CBF on a uniform neutralizing background and generalize to
non-integer dimensions.  The model Hamiltonian is given by
\begin{equation}
H = \int d\bfr \, \psi^\dagger (\bfr)\left( - \frac{\nabla^2}{2m} -\mu 
\right) \psi(\bfr) 
+ \frac{1}{2} \int \int d\bfr \,  d\bfr' \psi^\dagger (\bfr)  \psi^\dagger (\bfr')
V(|\bfr-\bfr'|) \psi (\bfr')  \psi (\bfr), 
\label{1}
\end{equation}
where $V(|\bfr-\bfr'|) = Q^2/|\bfr - \bfr'|$,
$m$ is the effective mass, and $Q$ is the effective
charge of the CBF.   $\psi(\bfr)$ is the boson field operator 
and $\mu$ is the chemical potential.
 $\hbar = c = 1$ is assumed for convenience.

 D-dimensional coupling strength is measured by the dimensionless 
parameter $r_s$ defined by $A r_s^D = (n a_B^D)^{-1}$.
A large $r_s$ means a long range of interaction and a small
density of particles.
 $A$ is the D-dimensional volume parameter 
given  by  $A = 2 \pi^{D/2} D^{-1} \Gamma(D/2)^{-1}$ and 
 $a_B$ is the effective Bohr radius of the system
defined by $a_B = 1/m Q^2$.

We  first  apply the Bogoliubov transformation at T=0. 
The field operator is written as
\begin{equation}
\psi(\bfr,t) = \psi_0 + {\tl \psi}(\bfr,t),
\label{2}
\end{equation}
where $\psi_0$ is a macroscopic order parameter representing
 the condensate. $\tl \psi$ describes the particles promoted out of
the condensate and can be expressed as a linear transformation
\begin{equation}
{\tl \psi}(\bfr,t) = \sum_k \left[ u_k(\bfr,t) a_k + v_k^\ast (\bfr,t)
a_k^\dagger \right],
\label{5}
\end{equation}
where $a_k(a_k^\dagger)$ is the bosonic quasi-particle operator. 
Also, we have $\mu=0$, and $n_0 = \psi_0^2$  for the uniform condensate.

A straightforward calculation produces the following results 
in  D-dimensions\cite{alex2,tosi2}
\begin{eqnarray} 
u_k^2 &=& \frac{1}{2} \left[ \eps_k^{-1} \left\{  
n_0 U_D(k) + \frac{k^2}{2m} \right\} +1 \right],
\label{13}
\\
v_k^2 &=& \frac{1}{2} \left[ \eps_k^{-1} \left\{ 
n_0 U_D(k) + \frac{k^2}{2m} \right\} -1 \right],
\label{15}
\end{eqnarray}
where
\begin{equation}
\eps_k = \left\{ \frac{n_0 k^2 U_D(k)}{m}
 + \left( \frac{k^2}{2m} \right)^2 \right\}^{1/2}.
\label{17}
\end{equation}
Here,  $\eps_k$ is the energy spectrum of the Bogoliubov transformation.
Note that $u_k^2 - v_k^2 = 1$. 
 $U_D(k)$ is the Fourier transformation 
of the  $D$-dimensional coupling  $V(r)$ and  
 given by $U_D(k) = D A Q^2 / k^{D-1}$.
Note that  $U_3(k) = 4\pi Q^2/k^2$, and $ U_2(k) = 2 \pi Q^2/k$
as expected.

\section{condensate fraction at finite temperatures and
non-integer dimensions}

The extension of the theory to finite temperatures is effected by
the Bose distribution function, 
$f_k = 1/\left\{ \exp(\beta \eps_k) - 1 \right\}$.
The condensate density at nonzero temperature in D-dimensions
is given by
\begin{eqnarray} 
n_0 &=& n- <\psi'^\dagger (\bfr) \psi'(\bfr)>
\nonumber \\
&=& n - \sum_{\bfk\neq 0} \left[ v_k^2 + f_k (u_k^2 + v_k^2) \right].
\label{19}
\end{eqnarray}

Substituting Eqs. (\ref{13}), (\ref{15}), and (\ref{17}) 
into Eq. (\ref{19}), we obtain the condensate fraction 
in D-dimensions  as follows
\begin{eqnarray}
1-\frac{n_0}{n} &=& \frac{1}{n}\sum_{\bfk\neq 0}
\left[ v_k^2 + f_k (u_k^2 + v_k^2)  \right]
\nonumber \\
&=& \frac{1}{2n}\int_0^\infty dk \frac{DA k^{D-1}}{(2\pi)^D} 
\left[ \frac{n_0 U_D(k) + \frac{k^2}{2m}}
{\sqrt{\frac{n_0 k^2 U_D(k)}{m} + \left( \frac{k^2}{2m} \right)^2}} 
\left(1+\frac{2}{e^{\beta\eps_k} -1} \right) -1 \right].
\label{21}
\end{eqnarray}

Eq. (\ref{21}) can be converted  into a compact
 form after some mathematical steps,
\begin{equation}
\left( 1-\frac{n_0}{n}\right) \left( \frac{n_0}{n} \right)^{-D/(D+1)}
= F(D) I(T,D).
\label{23}
\end{equation}
 $F(D)$ is the coefficient which is defined by 
\begin{equation}
F(D) = \frac{3}{D+1} \, \frac{ 2^{(-D^2+3D+2)/(D+1)}}{D^{(D+2)/(D+1)}}
\, \frac{r_s^{D/(D+1)}}{\Gamma^2\left( D/2 \right)}.
\label{25}
\end{equation}
$I(D,T)$ is the integral which is defined by
\begin{equation}
I(D,T) = \int_0^\infty dx \, x^{-(D-2)(D-3)/D(D+1)}
\left[ \frac{p(x)}{q(x)} 
\left\{ 1+ \frac{2}{e^{C(D,T) g(x) q(x)} -1}\right\} - 2 x^{6/D}\right],
\label{27}
\end{equation}
where
\begin{eqnarray}
p(x) &=& 1 + 2 x^{12/D},
\label{29}
\\
q(x) &=& \sqrt{1 + x^{12/D}},
\label{31}
\end{eqnarray}
\begin{equation}
C(D,T)=\left( \frac{D n_0}{n}\right)^{2/(D+1)}
\frac{2^{(3-D)/(D+1)}}{ r_s^{2D/(D+1)}} \frac{Q^2}{a_B k_B T},
\label{33}
\end{equation}
and 
\begin{equation}
g(x) = x^{6(3-D)/D(D+1)}.
\label{35}
\end{equation}
The temperature dependence is enclosed in the function $C(D,T)$.
The integral $I(D,T)$ is performed numerically.

The condensate fraction in D-dimensions is plotted as a function of 
temperature in FIG. 1.
The unit of the temperature is $Q^2/a_B k_B$.
The three coupling strengths were chosen for the calculation:
(a) $r_s = 0.8$ for a  weak coupling and
(b) $r_s= 1.2$ for an intermediate coupling, and (c) $r_s=2.0$
for a strong coupling, respectively.
We see that the condensate fraction depends strongly on the interaction 
strength, $r_s$.  We also observe  that for all the  different values
 of the coupling strength, 
the condensate fractions have significant
nonzero values and are  relatively flat near  $T \sim 0K$ region. 

\section{superfluid  fraction at finite temperatures and
non-integer dimensions}

In order to compare the theory with  available 
experimental data on superconducting films,
 it is necessary to obtain the superfluid fraction
which is directly related to the penetration depth of superconducting
films.
The superfluid density of dilute Bose gas in $D$-dimensions
has been studied by Fisher and Hohenberg \cite{fish}.
It is given by a Landau quasi-particle formula based on the 
$D$-dimensional Bogoliubov energy spectrum $\eps_k$ in Eq. (\ref{17}) as
\begin{equation}
\frac{n_s}{n} = 1 - \frac{\beta}{2 n m }
\sum_{\bfk\neq 0}  k^2 \frac{e^{\beta \eps_k}}{(e^{\eps_k}-1)^2}.
\label{51}
\end{equation}

Substitution of Eq. (\ref{17}) into the  Eq. (\ref{51}) gives an
explicit expression for the D-dimensional superfluid fraction,
\begin{equation}
\frac{n_s}{n} = 1 - H(D,T) \int_0^{\infty} dx \, x^{D+1}
 \frac{e^{B(D,T)\sqrt{x^{3-D}+x^4}}}{\{e^{B(D,T)\sqrt{x^{3-D} + x^4}}-1\}^2}.
\label{53}
\end{equation}
$B(D,T)$ and $H(D,T)$ are defined by the following expressions:
\begin{equation}
B(D,T)= 2^{(3-D)/(D+1)} 
\left(\frac{D n_0 r_s}{n} \right)^{2/(D+1)} \frac{Q^2}{a_B k_B T},
\label{55}
\end{equation}
and
\begin{equation}
H(D,T)=  \frac{A^2  D^{(2D+1)/(D+1)}}{(2\pi)^D} 
\left( \frac{ 4 n_0 r_s}{n}\right)^{D/(D+1)} B(D,T).
\label{57}
\end{equation}
The condensate fraction $n_0/n$  of Eq. (\ref{23}) is substituted into
 Eqs. (\ref{55}) and (\ref{57}) for the calculation of the 
superfluid fraction $n_s/n$.

The $D$-dimensional superfluid fraction is obtained from
 Eq. (\ref{53}), and plotted as a function of temperature for three
different values of the coupling strength in FIG. 2. 
The same $r_s$ values as in FIG. 1 were chosen for the comparison.

We find the basic structure between the condensate fraction in FIG. 1 
and the superfluid fraction in FIG. 2 is similar for any coupling
strength, $r_s$.
Therefore, it is clear that condensate density gives a strong and 
useful hint for the superfluid density for the CBF system.

\section{superfluid density of superconducting films}

Measurements on  the temperature dependent superfluid density
of superconducting films 
$La_{2-x} Sr_x CuO_4$  and $Mo_{77}Ge_{23}$
drew considerable interests\cite{andr,page6545,page641,lemb4,turn},
in connection with the puzzling splitting behaviors shown in FIG. 3.  
In FIG. 3 we note that $1/\lambda^2$ is proportional to the
superfluid density \cite{andr,page6545,page641,lemb4},
and inverse of the sheet inductance, $1/L(T)$, is also proportional 
to the areal superfluid density, $n_s d$\cite{turn}.

We observe that the two figures in FIG. 3 exhibit the same 
generic splitting behavior for samples with various thickness.
Considering differences in the detailed physical properties between
LSCO and MoGe films, such qualitative similarity is quite 
surprising and, thus, it strongly implies that 
the generic behavior stems from some hidden common properties.

There have  been several theoretical efforts to explain the general
 features of the superfluid density of superconducting films.
It has been shown that the simple BCS and the d-wave symmetry just
lead into a quadratic dependence of 
$1/\lambda^{2}(T)$\cite{page6545,page641,alex1}
and, thus, can not explain the generic splitting
behavior. Effect of thermal fluctuations is shown to explain
the relative flat behavior of curves at low temperatures\cite{alex1}.
But it could not also explain the above mentioned behavior.

Since the superconducting mechanisms of LSCO is believed to be basically
 different from that of MoGe, any theory which explains the behavior
 should not be based on the microscopic details of the superconducting 
mechanisms of the two materials.
 Instead, it should mainly reflect the common geometric nature
of superconducting films.

It has been shown that condensation of bound pairs can be used to 
explain the superconductivity with proper scaling\cite{lemb,scha}.
In the previous section, we have shown that the superfluid density,
which originates from the charged boson condensation in fractal dimensions,
$2<D<3$, exactly duplicates the observed experimental data from
the superconducting films. 
Therefore, we believe that the theoretical results in FIG. 2 
are the natural explanation of the experimental observations in FIG. 3.
  
We note also that the general features of the superfluid density 
curves of the superconducting films are qualitatively similar to 
those of liquid helium-4 in films and porous
media\cite{Brew,Bret,Fino,Steele}.

\section{Conclusions}

In this paper, we have studied  the condensate of charged boson fluid  
at finite  temperatures in non-integer dimensions between 2 and 3.
The condensate  and superfluid fraction are 
 obtained as  functions of temperature and dimensions 
at various values of the coupling strength.

We have shown that the generic splitting behavior of the superfluid 
density universal for superconducting films may originate from the
geometric nature of the films and detailed nature of the superconducting
mechanism plays a much less prominent role than generally believed.

\acknowledgments
This work was supported by the Korea Science and Engineering Foundation
through a special program and
the Center for Strongly Correlated Material Research(SNU).


\begin{figure}
\caption{The condensate fraction of the charged bosons in 
non-integer dimensions. 
From top  to bottom  $D=3.0, D=2.8, D=2.6, D=2.4$
  (a) When $r_s = 0.8$, (b) when $r_s = 1.2$, 
and (c) when $r_s = 2.0$.  
The unit of the temperature is $Q^2/a_B k_B$.}
\end{figure}
\begin{figure}
\caption{The superfluid fraction of the charged bosons in
non-integer dimensions. 
From top to bottom $D=3.0, D=2.8, D=2.6, D=2.4$
  (a) When $r_s = 0.8$, (b) when $r_s = 1.2$, 
and (c) when $r_s = 2.0$.  
The unit of the temperature is $Q^2/a_B k_B$.}
\end{figure}
\begin{figure}  
\caption{(a)  Superfluid fraction 
of $La_{2-x} Sr_x Cu O_4$ films measured by K. M. Paget {\it et al.}[9] 
The transition temperatures are  between $20K-30.5K$, 
and film thickness are between $5000\AA-900\AA$. 
The film thickness decreases from top to bottom.
(b) Superfluid fraction of the $Mo_{77}Ge_{23}$ 
superconducting films measured by S. J. Turneaure {\it et al.}[12]   
 The transition temperatures are between $2.999K-7.050K$, 
and film thickness are between $500\AA-20\AA$ from top to bottom. }
\end{figure}
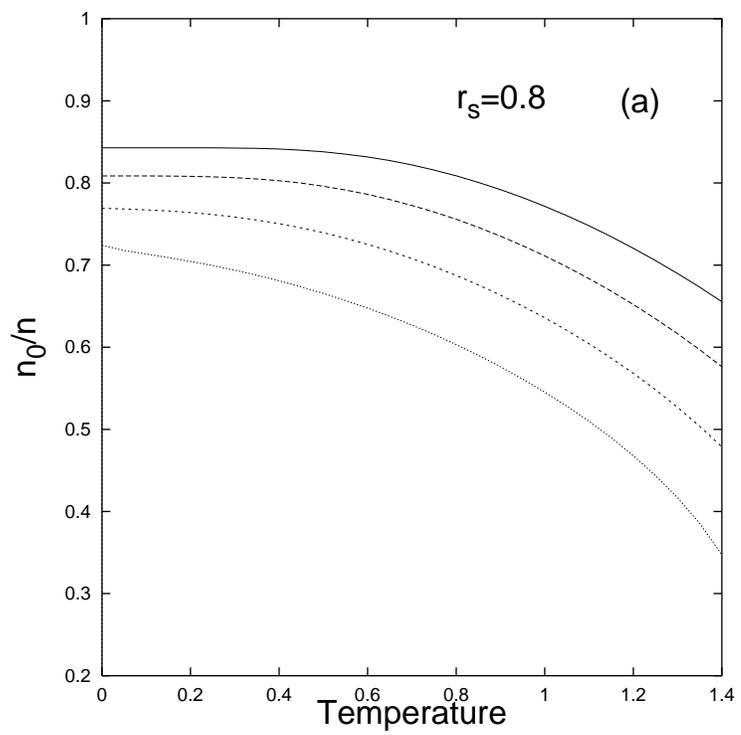
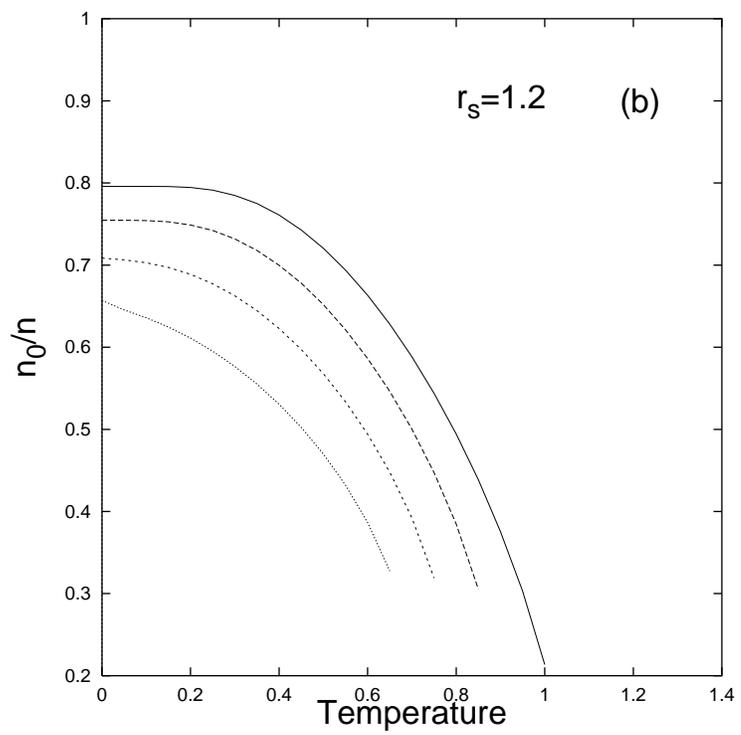
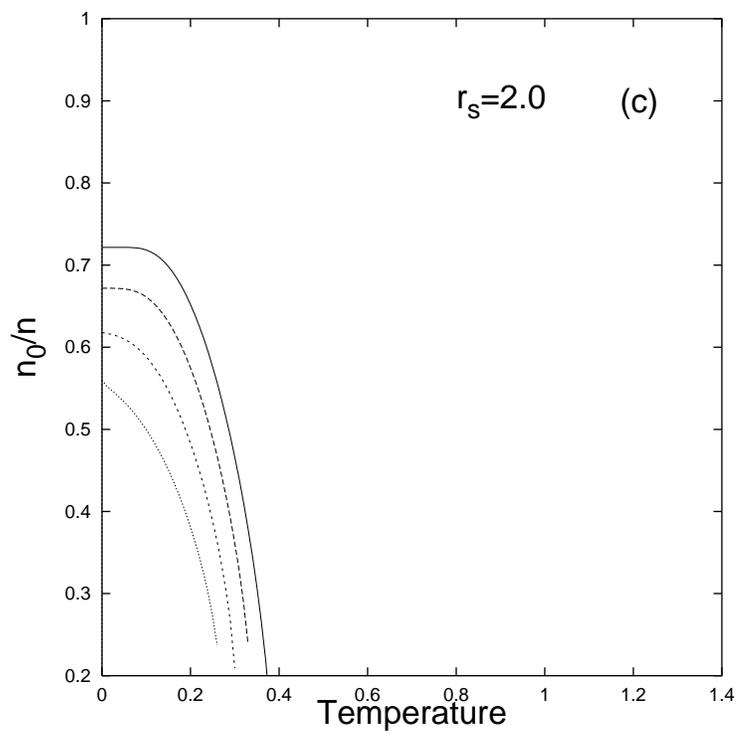
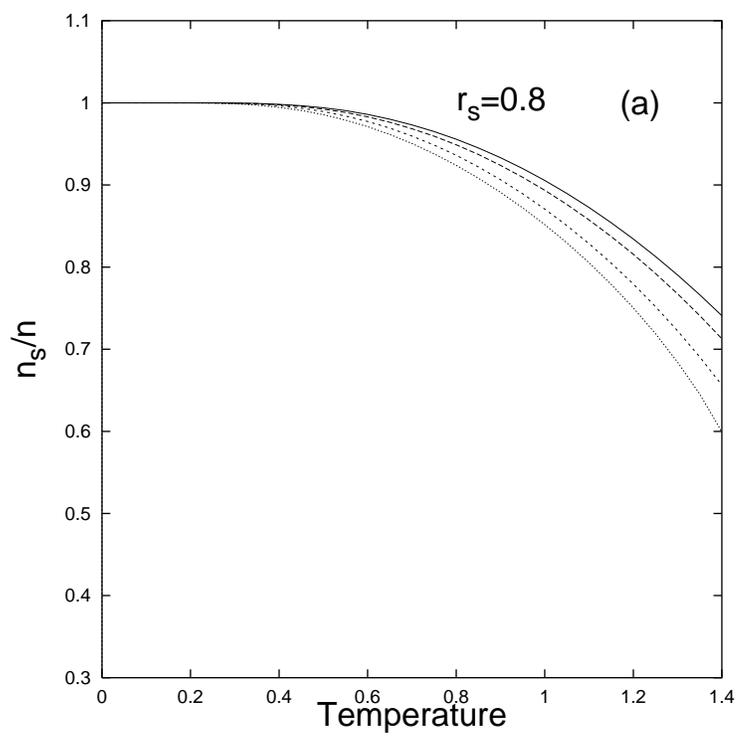
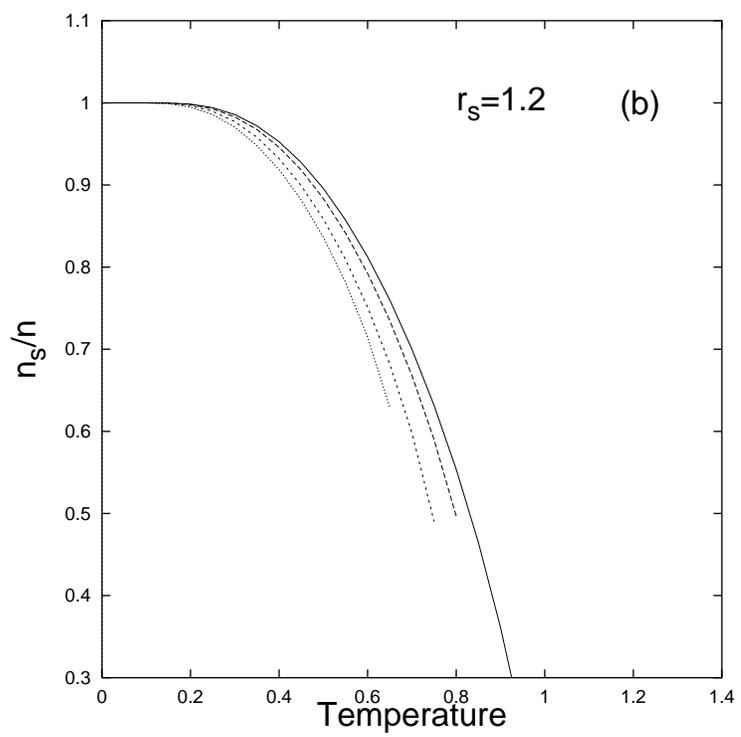
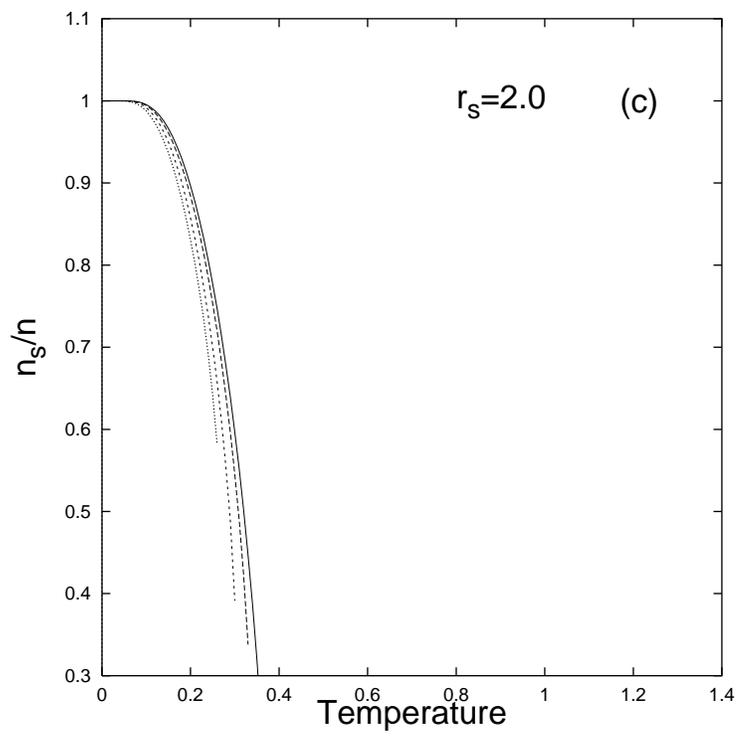
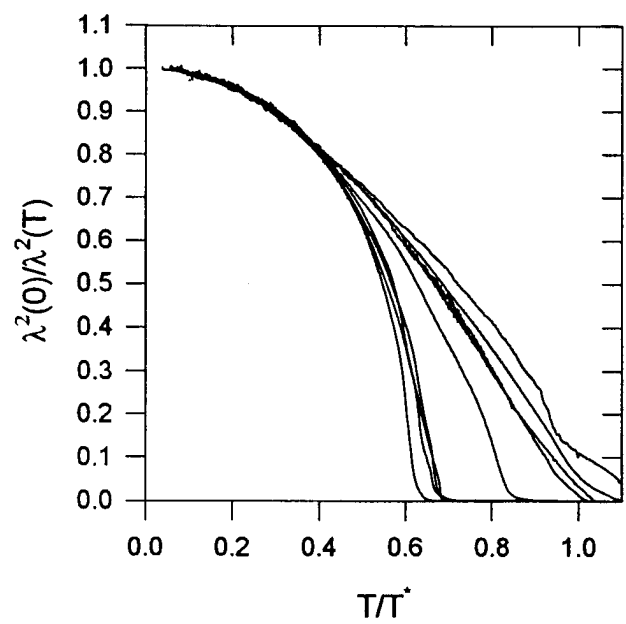
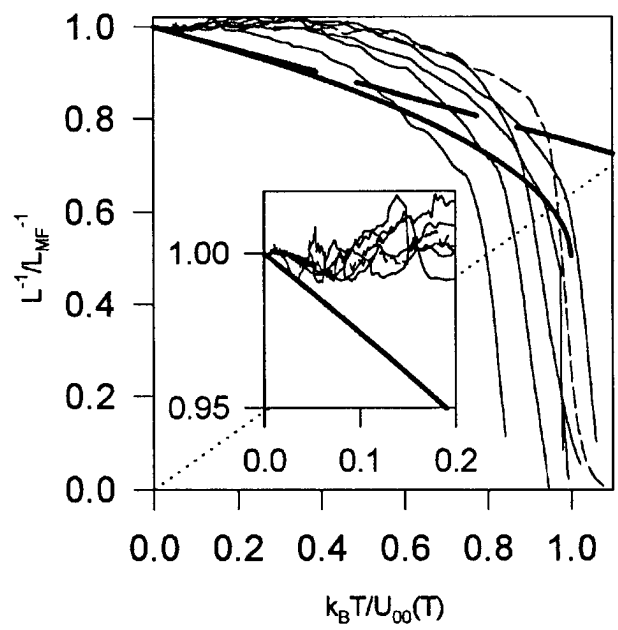
\end{document}